\documentclass[]{aastex62}
\graphicspath{{../images/}}
\usepackage{amsmath}

\begin{document}
	\title{Ethane in Titan's Stratosphere from Cassini CIRS Far- and Mid-Infrared Spectra}
	\shorttitle{Titan's Ethane}
	\shortauthors{Lombardo et al.}
	
	\correspondingauthor{Nicholas A Lombardo}
	\email{nicholas.lombardo@nasa.gov}
	
	\author{Nicholas A Lombardo}
	\affiliation{Center for Space Science and Technology, University of Maryland, Baltimore County, 1000 Hilltop Circle, Baltimore, MD, USA}
	\affiliation{Goddard Space Flight Center, 8800 Greenbelt Road, Greenbelt, MD, 20770, USA}
	
	\author{Conor A Nixon}
	\affiliation{Goddard Space Flight Center, 8800 Greenbelt Road, Greenbelt, MD, 20770, USA}
	
	\author{Melody Sylvestre}
	\affiliation{School of Earth Sciences, University of Bristol, Wills Memorial Building, Queens Road, Bristol, BS8 1RJ, UK}
	
	\author{Donald E Jennings}
	\affiliation{Goddard Space Flight Center, 8800 Greenbelt Road, Greenbelt, MD, 20770, USA}
	
	\author{Nicholas Teanby}
	\affiliation{School of Earth Sciences, University of Bristol, Wills Memorial Building, Queens Road, Bristol, BS8 1RJ, UK}
	
	\author{Patrick J G Irwin}
	\affiliation{Atmospheric, Oceanic and Planetary Physics, Clarendon Laboratory, University of Oxford, Parks Road, Oxford OX1 3PU, UK}
	
	\author{F Michael Flasar}
	\affiliation{Goddard Space Flight Center, 8800 Greenbelt Road, Greenbelt, MD, 20770, USA}
	
\begin{abstract}
	The Cassini Composite Infrared Spectrometer (CIRS) observed thermal emission in the far- and mid-infrared (from 10 cm$^{-1}$ to 1500 cm$^{-1}$), enabling spatiotemporal studies of ethane on Titan across the span of the Cassini mission from 2004 through 2017.  Many previous measurements of ethane on Titan have relied on modeling the molecule's mid-infrared $\nu_{12}$ band, centered on 822 cm$^{-1}$.  Other bands of ethane at shorter and longer wavelengths were seen, but have not been modeled to measure ethane abundance.
	
	Spectral line lists of the far-infrared $\nu_{4}$ torsional band at 289 cm$^{-1}$ and the mid-infrared $\nu_{8}$ band centered ay 1468 cm$^{-1}$ have recently been studied in the laboratory.  We model CIRS observations of each of these bands (along with the $\nu_{12}$ band) separately and compare retrieved mixing ratios from each spectral region.
	
	Nadir observations of of the $\nu_{4}$ band probe the low stratosphere below 100 km.  Our equatorial measurements at 289 cm$^{-1}$ show an abundance of (1.0$\pm$0.4) $\times$10$^{-5}$ at 88 km, from 2007 to 2017. This mixing ratio is consistent with measurements at higher altitudes, in contrast to the depletion that many photochemical models predict.
	
	Measurements from the $\nu_{12}$ and $\nu_{8}$ bands are comparable to each other, with the $\nu_{12}$ band probing an altitude range that extends deeper in the atmosphere.  We suggest future studies of planetary atmospheres may observe the $\nu_{8}$ band, enabling shorter wavelength studies of ethane. There may also be an advantage to observing both the ethane $\nu_{8}$ band and nearby methane $\nu_{4}$ band in the same spectral window.  
\end{abstract}

\keywords{planets and satellites: atmospheres --- planets and satellites: individual (Titan) --- infrared: planetary systems }

	\section{Introduction}
	
	The largest moon of Saturn, Titan, is unique among moons in the solar system for having a substantial atmosphere.  The two primary constituents are molecular nitrogen (N$_{2}$, comprising between 95\% and 99\% by volume), and methane (CH$_{4}$, 1.1\% - 1.5\% as measured in \cite{lellouch:14}).  Photodissociation of the two primary species give rise to molecular hydrogen (H$_{2}$, 0.1\% - 0.3\%, \cite{niemann:2010, cui:09}), carbon monoxide (CO, 50 ppm uniformly mixed, \cite{rengel:14}), ethane (C$_{2}$H$_{6}$, about 10 ppm in the lower stratosphere, from \cite{vinatier:2015}) and many larger hydrocarbons and nitriles.
	
	The first band of ethane detected on Titan was the $\nu_{12}$ band centered at 822 cm$^{-1}$ using the UCSD-University of Minnesota 60-inch (152-cm) telescope located at the Mount Lemmon Observing Facility by \cite{gillett:73} - though it was not definitively attributed to ethane until \cite{danielson:73}.  Using spatially resolved observations from the IRIS instrument on the Voyager 1 spacecraft, which made a flyby of Titan in 1981, \cite{coustenis:1989} searched for vertical and longitudinal variations in Titan's stratosphere, and concluded that the molecule was well mixed with a volume mixing ratio (VMR) of 1.3$^{+0.5}_{-0.7}\times$10$^{-5}$, consistent with today's measurements.  High spatial resolution observations from the Cassini spacecraft enabled the study of seasonal, meridional (latitudinal), and vertical variations in the abundance of ethane, which have been included in photochemical and dynamical models of Titan.  RADAR observations and theoretical modeling of the Ontario Lacus and Ligeia Mare lakes on Titan have shown their composition to be up to 40\% ethane, however the molecule was not detected in Punga Mare or Buffin Sinus (\cite{mastro:17, mastro:18}, see also \cite{cordier:09, cordier:12}).  In the stratosphere, between 100 km and 500 km, ethane has a roughly constant abundance of 10$^{-5}$, as measured by the Composite Infrared Spectrometer (CIRS) \citep{vinatier:2007, vinatier:2015}.  In situ measurements of Titan's thermosphere near 1000 km by the Ion and Neutral Mass Spectrometer (INMS) also on Cassini have shown ethane to have an abundance of about 4.5$\times$10$^{-5}$ \citep{magee:2009}.  The abundance of ethane between the surface and 100 km is not well known.  At the time of this study, accurate measurements of ethane from the Huygens Gas Chromatograph and Mass Spectrometer (GCMS) have not been reported, and only approximate abundances with large uncertainties are given in \cite{niemann:2010}.  The team reports an ethane abundance at 88 km between 10$^{-6}$ and 1.2$\times$10$^{-5}$, over an order of magnitude in range.  Radiative transfer modeling of ground-based and CIRS spectra have also been unable to probe below about 100 km due to the high opacity of available spectral bands.  Even if nadir observations are used, thermal emission from this band generally originates from altitudes above 100 km \citep{coustenis:10}.
	
	Shortly after the arrival of the Cassini spacecraft to the Saturn system, \cite{griffith:2006} discovered ethane clouds at latitudes poleward of 50$^\circ$N at altitudes between 30 km and 50 km.  Discovered while the north pole was in winter, the clouds evolved and eventually dissipated as Titan entered equinox \citep{lemouelic:12}.  
	
	In the mid-infrared, between about 600 and 3000 cm$^{-1}$ (16.7 and 3.3$\mu$m), ethane has several infrared-active rovibrational bands including the $\nu_{1}$, $\nu_{5}$, $\nu_{7}$ and $\nu_{10}$ bands between 2700 cm$^{-1}$ and 3100 cm$^{-1}$, the $\nu_{2}$, $\nu_{8}$, $\nu_{6}$, and $\nu_{11}$ bands between 1350 cm$^{-1}$ and 1600 cm$^{-1}$, and the $\nu_{12}$ band centered at 822 cm$^{-1}$.  Of these bands, the strong $\nu_{12}$ band has been used the most to model Titan's atmospheric spectra and extract ethane abundance between 100 km and 500 km \citep{vinatier:2007, vinatier:2015}.
	
	Previous attempts to retrieve ethane abundance from the $\nu_{6}$ and $\nu_{8}$ bands between 1350 cm$^{-1}$ and  1500 cm$^{-1}$ have proved to be difficult. In particular, \cite{coustenis:10} found it difficult to extract precise abundances of ethane from this region potentially due to a lack of C$_{3}$H$_{8}$ line data at the time and inaccurate line data for C$_{2}$H$_{6}$.  Since then, new line data for C$_{2}$H$_{6}$ and C$_{3}$H$_{8}$ have been made available, allowing our analysis to proceed \citep{dilauro:2012, sung:propane}
	
	In the far-infrared, ethane has a single mode, the $\nu_{4}$ torsional band centered at 289 cm$^{-1}$.  Until recently, spectral line lists for this band were not available, and thus retrievals utilizing this band were not possible.  However, lab work presented in \cite{Moazzen-Ahmadi:15} has enabled the creation of a spectral line list for this band of ethane.  The $\nu_{4}$ band is less intense than the $\nu_{12}$ band, thus nadir observations are able to probe deeper through Titan's atmosphere before optical saturation.  Therefore, deeper soundings of ethane abundance on Titan are now possible by modeling the $\nu_{4}$ band seen in nadir observations of Titan's disk, described in this paper for the first time.
	
	\section{Methods}
	
	The Composite Infrared Spectrometer (CIRS) was a Fourier Transform Infrared Spectrometer on the Cassini spacecraft that explored the Saturn system from 2004 until its intentional deorbit into Saturn in September 2017.  CIRS had three focal planes, each sensitive to a different region of the infrared spectrum.  Focal Plane 1 (FP1) was sensitive in the far-infrared from 10 cm$^{-1}$ - 600 cm$^{-1}$, FP3 was sensitive in the mid-infrared from 600 cm$^{-1}$ - 1100 cm$^{-1}$, and FP4 was sensitive in the mid-infrared from 1100 cm$^{-1}$ - 1500 cm$^{-1}$ \citep{jennings:17}.
	
	\subsection{ FIR Dataset} 
	FP1 comprised a single detector, with a field-of-view of about 4 mrad.  Nadir observations in this study were performed at a spectral resolution of 0.5 cm$^{-1}$ at emission angles between 45$^{\circ}$ and 85$^{\circ}$, averaged over the footprint of the detector, at a range from 1.5$\times$10$^{5}$ to 3.5$\times$10$^{5}$ km from Titan.  At these distances, the size of the detector footprint on the surface of Titan varies between 600 and 1400 km.  Data observed at high emission angles will have traveled a longer path length though the atmosphere, thus increasing the signal to noise of the $\nu_{4}$ ethane emission.  Though data observed at lower emission angles will have traveled a shorter path length, the ethane emission will have originated from the same source altitude.
	
\begin{figure}[h]
		\includegraphics[width = \columnwidth]{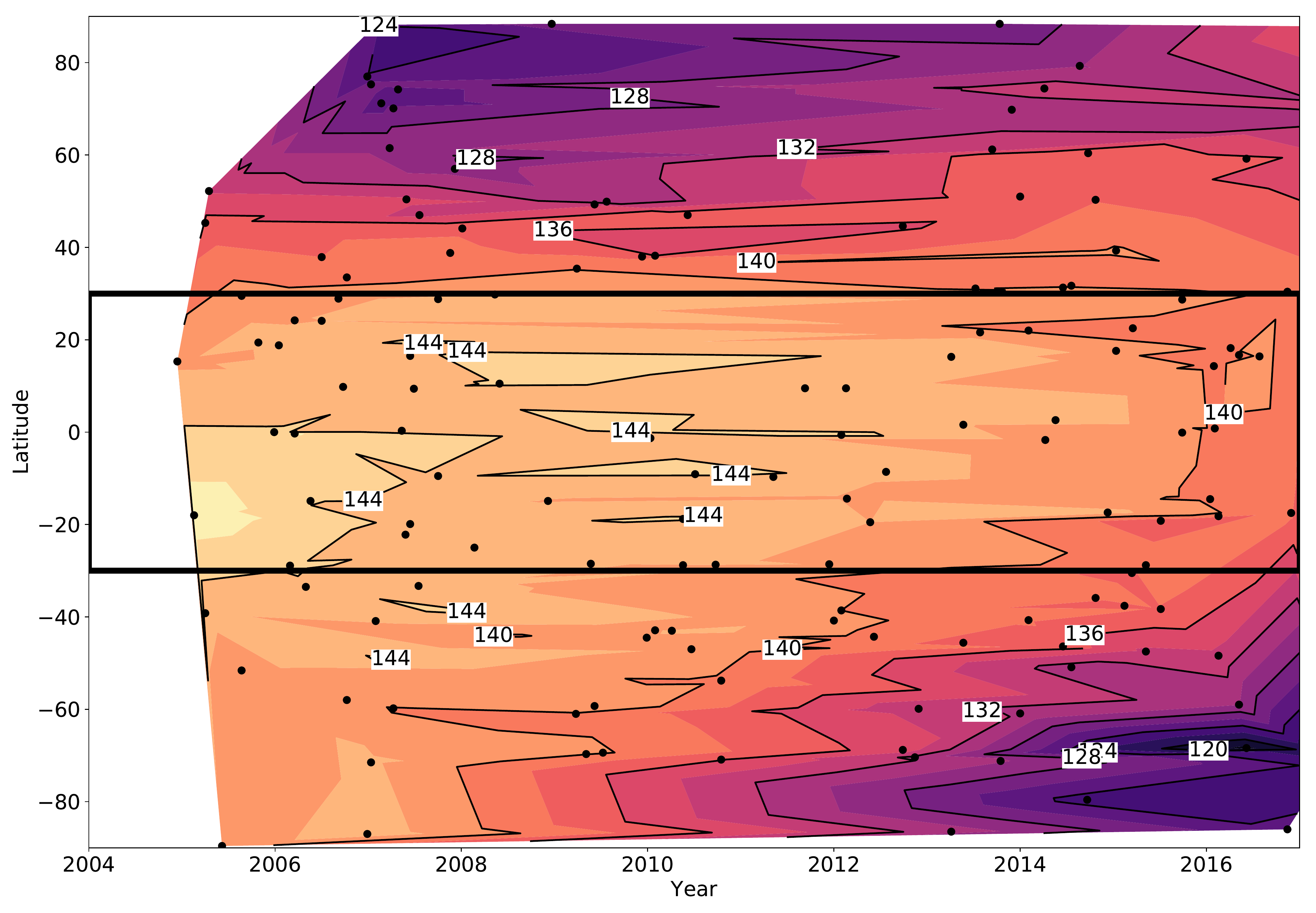}

		\caption{Temperatures at 15 mbar derived from Sylvestre, et al., submitted.  Black dots are FP1 nadir observations used to extract temperature.  The color map is interpolated from temperatures of individual observations.  The boundary of the averaging bin is shown as the black box. \label{fig:temperaturemap}}
	\end{figure}

	To increase the signal-to-noise of trace gases in the spectra we modeled, we combined observations between 30$^{\circ}$S and 30$^{\circ}$N, from 2007 through 2017.  These latitude bounds were chosen because, as shown in Fig.~\ref{fig:temperaturemap}, Titan's temperature at the altitude where our model is sensitive to (about 88 km) is similar over the duration of the mission.  This time range was chosen as it includes most of the data taken by CIRS in this latitude region, while also excluding a period of time in 2007 where a large noise feature was present in the FP1 spectra.  This average includes data from 6624 spectra.
	
	A map of the temperatures at 15 mbar (about 88 km, where our model is sensitive to) is shown in Fig.~\ref{fig:temperaturemap}.  Temperatures are from \cite{syl:19}, and are calculated by modeling the CH$_{4}$ rotational lines in FP1.
	
	\subsection{MIR Dataset}
	
	FP3 and FP4 were both linear arrays of 10 detectors, with a field-of-view of about 0.27 mrad per detector.  The smaller field-of-views of FP3 and FP4 allowed for vertically resolved observations of Titan's limb and vertical profiles of molecular abundances to be measured.  In the observations used in this study, the centers of field-of-view of the FP3 and FP4 arrays were positioned normal to Titan's surface at two tangent altitudes of 125 km and 225 km.  Limb observations in this study were performed at a spectral resolution of 0.5 cm$^{-1}$.  Spectra were vertically binned into non-overlapping 50 km bins, from 100 km - 400 km.  To enable a proper comparison to our FP1 measurements, we make use of the same time-latitude averaging scheme as our FP1 analysis.
	
	In FP3, we model the C$_{2}$H$_{6}$ $\nu_{12}$ band which has been extensively modeled in previous literature.  In FP4 we model the $\nu_{8}$ band centered at 1468 cm $^{-1}$.  Also present in FP4 is the $\nu_{6}$ band centered at 1379 cm$^{-1}$.  We included this band in our modeling, however results derived from this band were noisy and unreliable, due to the presence of stronger methane and propane emissions.
	
	A summary of the bands modeled in this study is shown in Table \ref{tab:bands}.
	
	\begin{table}[h]
		\centering
		\label{tab:bands}
		\begin{tabular}{ccccc}
			Band 		  & Band Center   &		Wavenumbers Modeled				& Observation Type  &Altitude Sensitivity \\
			\hline
			$\nu_{4}$	&  		289 cm$^{-1}$&240 cm$^{-1}$	- 300 cm$^{-1}$  &  FP1 Nadir  &  88 km\\
			$\nu_{12}$ & 	822 cm$^{-1}$	&800 cm$^{-1}$ - 860 cm$^{-1}$    & FP3 Limb&  152 km - 380 km\\
			$\nu_{8}$ 	& 	1468 cm$^{-1}$	&1440 cm$^{-1}$ - 1480 cm$^{-1}$& FP4 Limb&  212 km - 373 km\\
		\end{tabular}
		\caption{Table of ethane bands modeled in this study, as well as a summary of the spectral bins used and altitude sensitivity of the retrievals.  The altitude sensitivity lists only the altitude of maximum contribution for each bin, and does not include the FWHM of the contribution function.  We do not make use of mid-infrared nadir observations, as previous studies have shown that nadir observations of the $\nu_{12}$ band only probe to about 100 km \citep{bampasidis:2012}.  \label{tab:bands}}
	\end{table}

	\subsection{Spectral Line Data}
	Spectral line data used in the modeling of all 3 focal planes were taken from the 2016 HITRAN database \citep{hitran:2016}, which includes several bands of ethane, including the $\nu_{12}$ band, $\nu_{4}$ band (first studied in \cite{Moazzen-Ahmadi:15}), and the $\nu_{6}$ and $\nu_{8}$ bands (originally reported in \cite{dilauro:2012}).  Also from HITRAN are line data for H$_{2}$O and CH$_{4}$.  Line data for C$_{2}$N$_{2}$, and C$_{3}$H$_{4}$, not available in HITRAN, were taken from the GEISA database \citep{geisa:2015}.
	Line data for propane is not available in HITRAN, so our modeling makes use of the pseudo-line list presented in \cite{sung:propane}.
	
	\subsection{Radiative Transfer Modeling}
	Spectral modeling was performed with the NEMESIS inverse radiative transfer code \citep{irwin:nemesis}.  NEMESIS operates on the method of optimal estimation, which involves the computation of a forward model and a retrieval process. Forward models were calculated with the correlated-k method described in \cite{lacis:1991}.  To recreate the instrumental line shape, the model includes a Hamming apodization of Full Width at Half Maximum of 0.475 cm $^{-1}$.   The retrieval process operates by varying user-defined \textit{a priori} profiles of chosen physical parameters and \textbf{assumed uncertainties} (such as temperature, aerosol abundance, and trace gas volume mixing ratios) to optimize the spectral fits.  It is important to define a realistic \textit{a priori} profile and uncertainty, as providing unrealistic conditions will yield unrealistic results.  For example, defining the \textit{a priori} error too small on a gas profile that is being retrieved can over-constrain the model, the reader is directed to \cite{irwin:nemesis} for more information. Optimization of the fit of the synthetic spectrum to the measured spectrum is performed by minimizing the cost-function, a parameter that includes the deviation of the retrieved profile from the \textit{a priori} estimate, and the quality of fit to the spectra (similar to a $\chi^{2}$ goodness of fit test).  NEMESIS has been extensively used to determine atmospheric abundances in the outer solar system using IR spectra, and application to Titan is described in \cite{teanby:2009}, \cite{cottini:2012}, \cite{syl:18}, and \cite{lombardo:2019} and references therein.  
	
	In FP1, we model the spectral window from 240 to 300 cm$^{-1}$.  A temperature profile derived from a weighted average of temperatures reported in \cite{syl:19} was used.  Trace gas features in this region include H$_{2}$O at 253 cm$^{-1}$, C$_{2}$N$_{2}$ centered at 233 cm$^{-1}$ with emission from weak lines noticeable at wavenumbers less than 250 cm$^{-1}$ and more prominent in the winter hemisphere, and C$_{2}$H$_{6}$ at 289 cm$^{-1}$.  Also contributing to the spectral radiance in this region is H$_{2}$ which contributes to the collision induced absorption and an aerosol haze emission that contributes to the continuum.  Over the small spectral window that we model, we assume that aerosol haze has a gray spectral dependence (or spectrally flat), consistent with aerosol spectra retrieved in \cite{anderson:2011}.  C$_{2}$H$_{6}$, C$_{2}$N$_{2}$, and the aerosol haze are allowed to vary with altitude in the retrieveal process, while H$_{2}$ is allowed to vary as a constant-with-altitude profile.  H$_{2}$O is set at a constant abundance of 2$\times$10$^{-10}$, consistent with \cite{cottini:2012}.  At slightly higher wavenumbers, C$_{3}$H$_{4}$ contributes strongly to the spectral radiance, however contribution at wavenumbers less than 300 cm$^{-1}$ is negligible.  Further discussion of modeling the continuum in this region is discussed in the following subsection.
	
\begin{figure}[h]
		\includegraphics[width=\columnwidth]{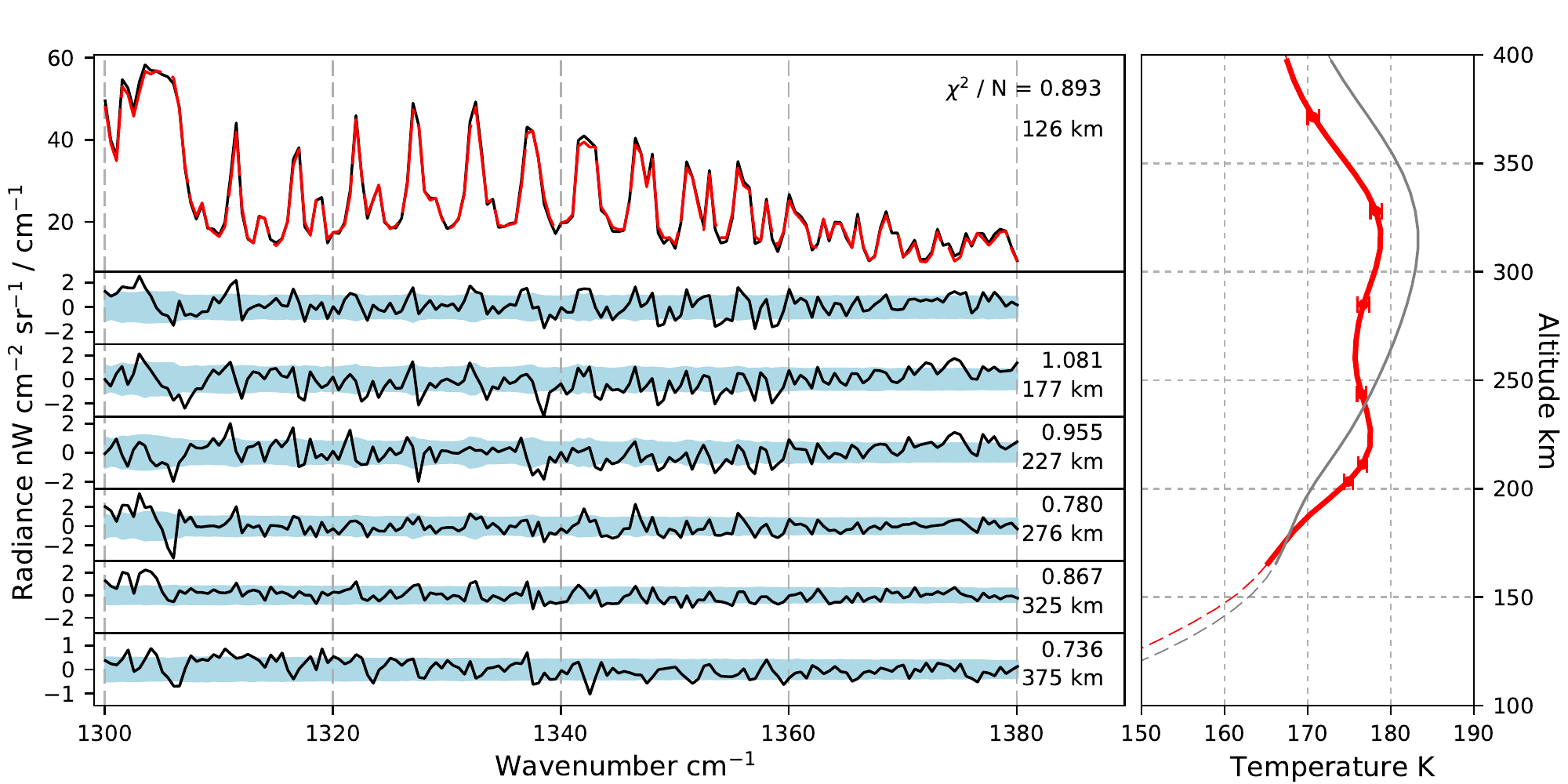}
		\caption{Model fits for the FP4 temperatures retrievals.  Left: The original data (black) compared to the synthetic spectrum (red dashed) for the lowest altitude bin (topmost plot).  The residuals (observed data minus synthetic spectra) are shown for all modeled spectra below.  Blue envelopes are the model error.  The values labeled on each spectrum are the modified $\chi^{2}$ value (top) and center altitude of each bin (bottom). Right: The retrieved temperature profile (red) and error bars shown with the \textit{a priori} (gray) profile used.  The \textit{a priori} is a weighted average of temperatures presented in \cite{achterberg:2014}. \label{fig:fp4temps}}
	\end{figure}

	For the MIR modeling, we first retrieve a mean temperature profile for the 30$^\circ$S - 30$^\circ$N latitude bin for the time range range covered by the MIR observations.  We model the 1300 - 1380 cm$^{-1}$ region of FP4, including the Q and R branches of the $\nu_{4}$ band of methane.  Our retrieval process involves setting a constant methane abundance of 1.1\% above the saturation altitude, as measured in \cite{lellouch:14} and allowing a temperature \textit{a priori} profile derived from a weighted average of temperatures presented in \cite{achterberg:2014} to vary.  An aerosol haze contributing to the continuum of the spectra is also allowed to vary with altitude in the retrieval process.  The modeled spectra and mean temperature profile used in our MIR modeling is shown in Fig. \ref{fig:fp4temps}.
	
	In FP3, we model the spectral window from 800 to 860 cm$^{-1}$.  This region is dominated by the $\nu_{12}$ band of ethane, however the $\nu_{8}$ band of propane (see \cite{nixon:propane}), centered at 869 cm$^{-1}$, may contribute to the continuum in the higher wavenumber end of this region.  In addition to these molecules, we also include a non-gray haze to model the continuum.  The opacity of the haze increases linearly with wavenumber over this region.
	
	In FP4, we model a spectral window from 1440 to 1480 cm$^{-1}$ including the $\nu_{8}$ band centered at 1468 cm$^{-1}$.  This window includes the $\nu_{17}$ and $\nu_{24}$ bands of propane, of which only the $\nu_{24}$ Q-branch at 1472 cm$^{-1}$ is noticeable (Fig. \ref{fig:fp4fit}).  We also include a non-gray haze to model the continuum in this region.  A gray haze does not accurately model the continuum in the higher wavenumbers of FP4, so we include a haze spectral response derived from \cite{vinatier:aerosol12}.
	
	\subsubsection{Unidentified Far-Infrared Continuum Feature}
\begin{figure}[h]
		\includegraphics[width=\columnwidth]{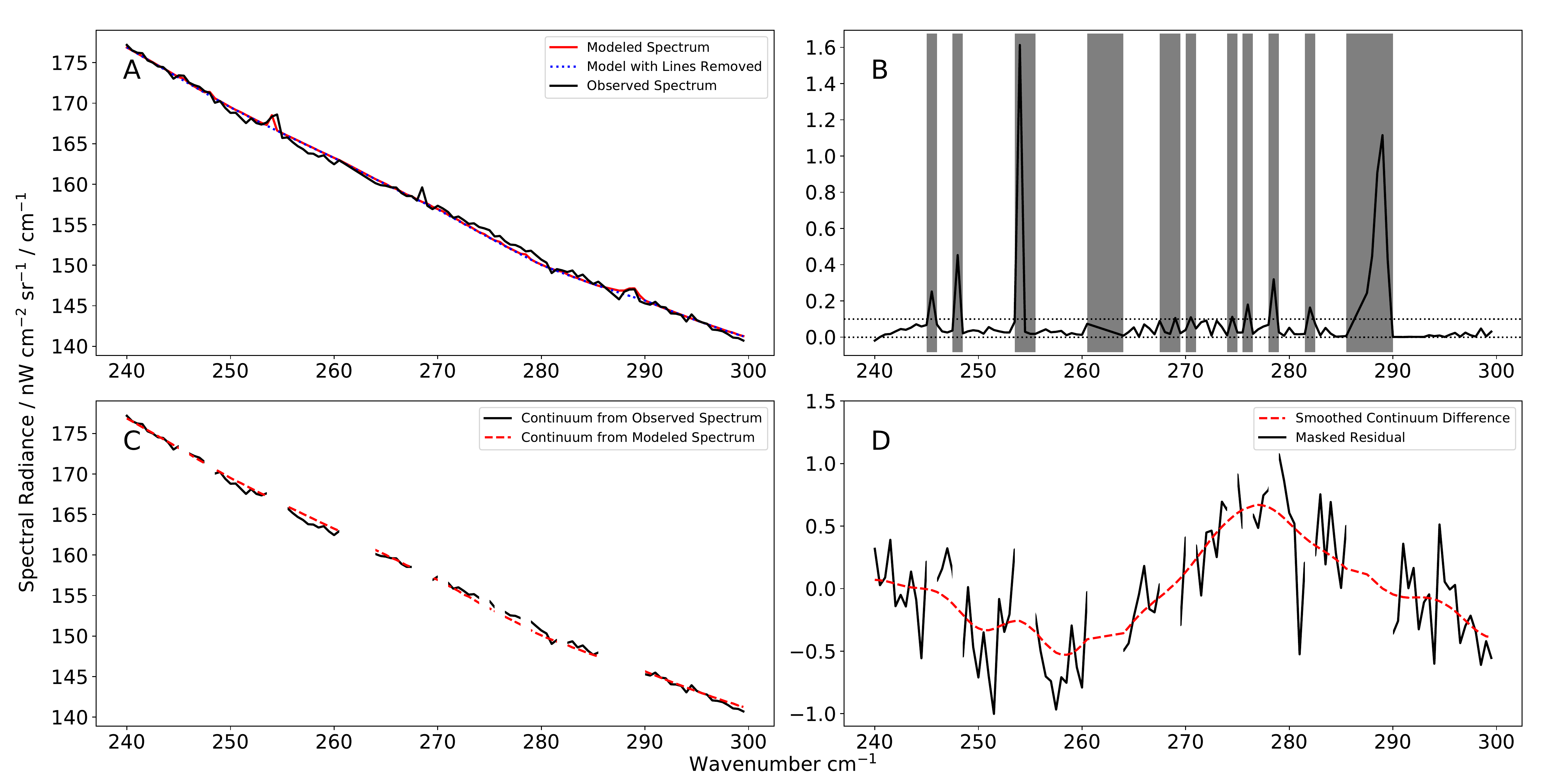}
		\caption{A - The observed spectrum (black), modeled spectrum (red), and modeled spectrum with molecular lines removed (black dotted).  The unmodelled continuum variation can be seen roughly between 270 cm$^{-1}$ and 290 cm$^{-1}$. B - The contribution from molecular lines, calculated by subtracting the model calculated with no molecular lines from the modeled spectrum.  Wavenumbers where the gas contribution is above 0.1 nW are masked to modify only the continuum, and are highlighted by gray boxes in the figure. C - The continuum (region where molecular contribution is minimal) from the model compared to the observations.  D - The difference between the observed and modeled continua (black).  The red dashed line is the smoothed and used as the spectral dependence for a second aerosol haze.  \label{fig:aerosol}}
	\end{figure}

	While only including spectral characteristics of H$_{2}$O, C$_{2}$N$_{2}$, C$_{2}$H$_{6}$, H$_{2}$, and a gray aerosol, we noticed a prominent emission feature in the residual of the synthetic spectrum (Fig. \ref{fig:aerosol}).  The feature is broad, spanning the region between 270 and 290 cm$^{-1}$, and appears in most bins.  To model this feature, we take an approach similar to \cite{teanby:13}. 
	
	First, we perform a retrieval as previously described, allowing the aforementioned parameters to vary.  Then, we remove spectral line data from H$_{2}$O, C$_{2}$N$_{2}$, and C$_{2}$H$_{6}$, and using the retrieved haze and H$_{2}$ abundance, run a forward model.  This allows us to examine only the continuum of the modeled spectrum, Fig.~\ref{fig:aerosol}A.  We then subtract this forward modeled continuum from the synthetic spectrum to identify the contribution from molecular lines, Fig.~\ref{fig:aerosol}B.  We then mask regions where this difference exceeds 0.1 nW cm$^{-2}$ sr$^{-1}$ / cm$^{-1}$, so that we identify wavenumbers that show only continuum.  The difference between the masked retrieved and masked forward modeled residuals is then smoothed and set as the extinction cross section for the haze representing the unidentified feature, Figs. \ref{fig:aerosol}C and D.  The cause of this feature could be a small-scale spectral structure in the haze that is only apparent in this high SNR large spectral average.
	
	\section{Results}
	\subsection{FP1}
\begin{figure}[h]
		\includegraphics[width = \columnwidth]{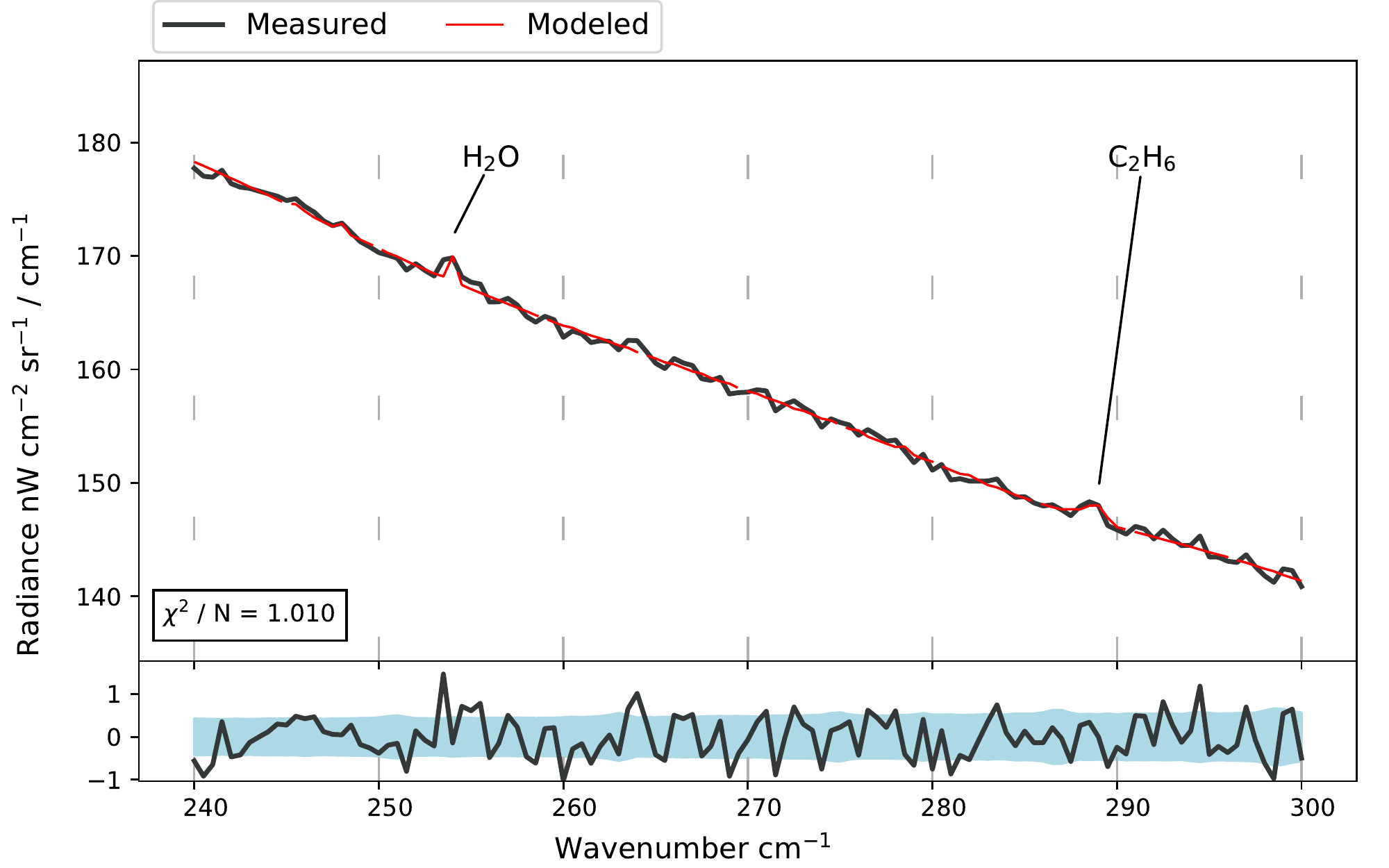}
		\caption{Upper panel: Synthetic spectrum (red) compared with the observed spectrum (black) of the $\nu_{4}$ band.  Lower panel: The residual of the fit (black), or the synthetic spectrum minus the observed spectrum, and 1 - $\sigma$ error envelope (light blue).  Where the residual is positive, the model is dimmer than the observation, and where the residual is negative, the model is brighter than the observation.  When the continuum is modeled with the previously described method, we are able to accurately model this spectral region.  The contribution from ethane to the spectrum can be seen as a small emission in the spectrum, which we can fit well.  \label{fig:fp1fit}}
	\end{figure}

	We compare the synthetic spectra to the observed spectra in Fig.~\ref{fig:fp1fit}.  The quality of the fit can be assessed with the $\chi^{2}$ / N value, which can be defined as 
	\begin{equation}
	\chi^{2} / N =  \dfrac{1}{N} \sum\limits_{\nu} \bigg( \dfrac{M(\nu) - O(\nu)}{\sigma(\nu)}\bigg) ^{2}
	\end{equation}
	
	where M is the modeled spectrum, O is the observed spectrum, $\sigma$ is the noise on the observed data, and N is the number of independent data points in the spectrum.  A synthetic spectrum with a $\chi^{2}$ / N of 1 is then the `best' possible fit to the observations.
	
	The C$_{2}$H$_{6}$ mixing ratio determined at 88 km (see Fig. \ref{fig:fp3_cf}) from the FP1 analysis is (1.0$\pm$0.4)$\times$10$^{-5}$, very close to our \textit{a priori} value of (1.0 $\pm$ 0.5)$\times$10$^{-5}$.  To check the sensitivity of the $\nu_{4}$ band to the \textit{a priori}, we also modeled the FP1 spectral window using two additional \textit{a priori} of (3.0$\pm$1.5)$\times$10$^{-5}$ and (5.0$\pm$2.5)$\times$10$^{-6}$.  The higher \textit{a priori} retrieved an ethane abundance of (1.2$\pm$0.4)$\times$10$^{-5}$, and the lower \textit{a priori} retrieved an abundance of (5.6$\pm$1.4)$\times$10$^{-6}$, both sensitive to the same altitude.  The abundances determined from each \textit{a priori} are within the model uncertainties, therefore we use the 10$^{-5}$ \textit{a priori} as it is in agreement with our $\chi^{2}$ analysis described in the Section 3.1.1.
	
	The contribution function, or the rate of change of spectral radiance with respect to the amount of the molecule included in the model, is a measure of where the model is sensitive for a certain molecule.  In Fig.~\ref{fig:fp3_cf}, we show the contribution function for ethane retrieved in the 30$^{\circ}$S to 30$^{\circ}$N bin, centered at 289 cm$^{-1}$.  The nadir contribution function obtains a maximum value at 13.1 mbar, or an altitude between 85.7 and 87.5 km.  The FWHM of the contribution function extends from 28 mbar to 3.7 mbar.

	\subsubsection{$\chi^{2}$ Analysis}
	
\begin{figure}[h]
		\centering
		\includegraphics[width = 0.5 \columnwidth]{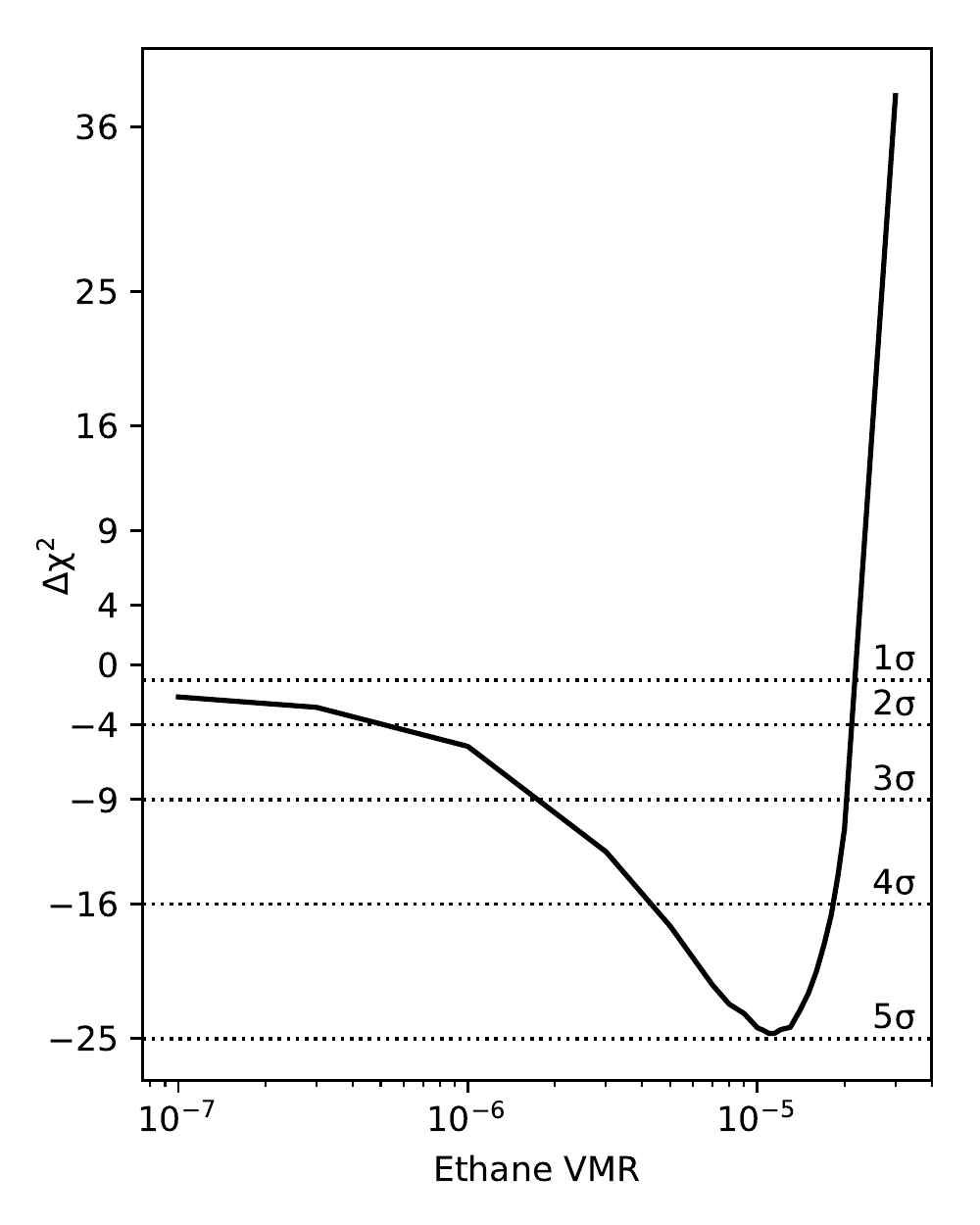}
		\caption{The $\Delta\chi^{2}$ values plotted against ethane abundance in the forward model.  The $\Delta\chi^{2}$ value achieves a minimum of -25 at 1.1$\times$10$^{-5}$, consistent with the retrieved measurement.  \label{fig:chisq}}
	\end{figure}

	In addition to the full retrieval already discussed, we performed a $\Delta\chi^{2}$ analysis.  In this analysis, similar to that presented in \cite{nixon:allene} and \cite{lombardo:2019}, we first perform a retrieval on the spectra including all molecules and aerosols previously described with the exception of ethane.  The $\chi^{2}$ of this retrieval will be referred to as $\chi^{2}_{0}$ (note that this is different from the $\chi^{2}$/N value previously described).  We then set the abundance profiles of these molecules to be the retrieved values, and run forward models with varying amounts of ethane added.  The ethane abundance profiles used in this method are set to be a constant value until  saturation, where they follow the saturation vapor pressure curves throughout the troposphere.  The $\chi^{2}$ value of the fit for each of the forward model runs will be referred to as $\chi^{2}_{m}$.  The $\Delta\chi^{2}$ value is then $\chi^{2}_{m}$ - $\chi^{2}_{0}$.  Where the $\Delta\chi^{2}$ achieves a minimum value, we claim as the most probable abundance of ethane, with a confidence of $\sqrt{\Delta\chi^{2}}$.  The 1 $\sigma$ confidence can be described as where $\Delta\chi^{2} - \Delta\chi_{minimum}^{2}$ = 1.  Figure \ref{fig:chisq} shows the plot of $\Delta\chi^{2}$ versus ethane abundance.  Using this method, we determine an ethane abundance of 1.1$^{+0.3}_{-0.2}$$\times$10$^{-5}$, very close to the abundance determined using a 1$\times$10$^{-5}$ \textit{a priori} profile and full retrieval.  The uncertainties on this value are the ethane VMRs such that $\Delta\chi^{2}_{minimum} - \Delta\chi^{2}_{vmr} = \pm 1$.  It is expected that the abundance retrieved using NEMESIS is sightly different from the abundance determined using this $\Delta\chi^{2}$ analysis since NEMESIS will vary the abundance of several molecules at once to optimize the spectral fit of the model to the data.  The measurement uncertainty reported on the retrieved abundance is slightly greater due to uncertainties in line position, line strength, temperature, and observation error that are included in the retrieval.  This $\Delta\chi^{2}$ method, however, does not include these uncertainties, hence the smaller error bars.
	
	\subsection{FP3}
	
\begin{figure}[h]
		\includegraphics[width = \columnwidth]{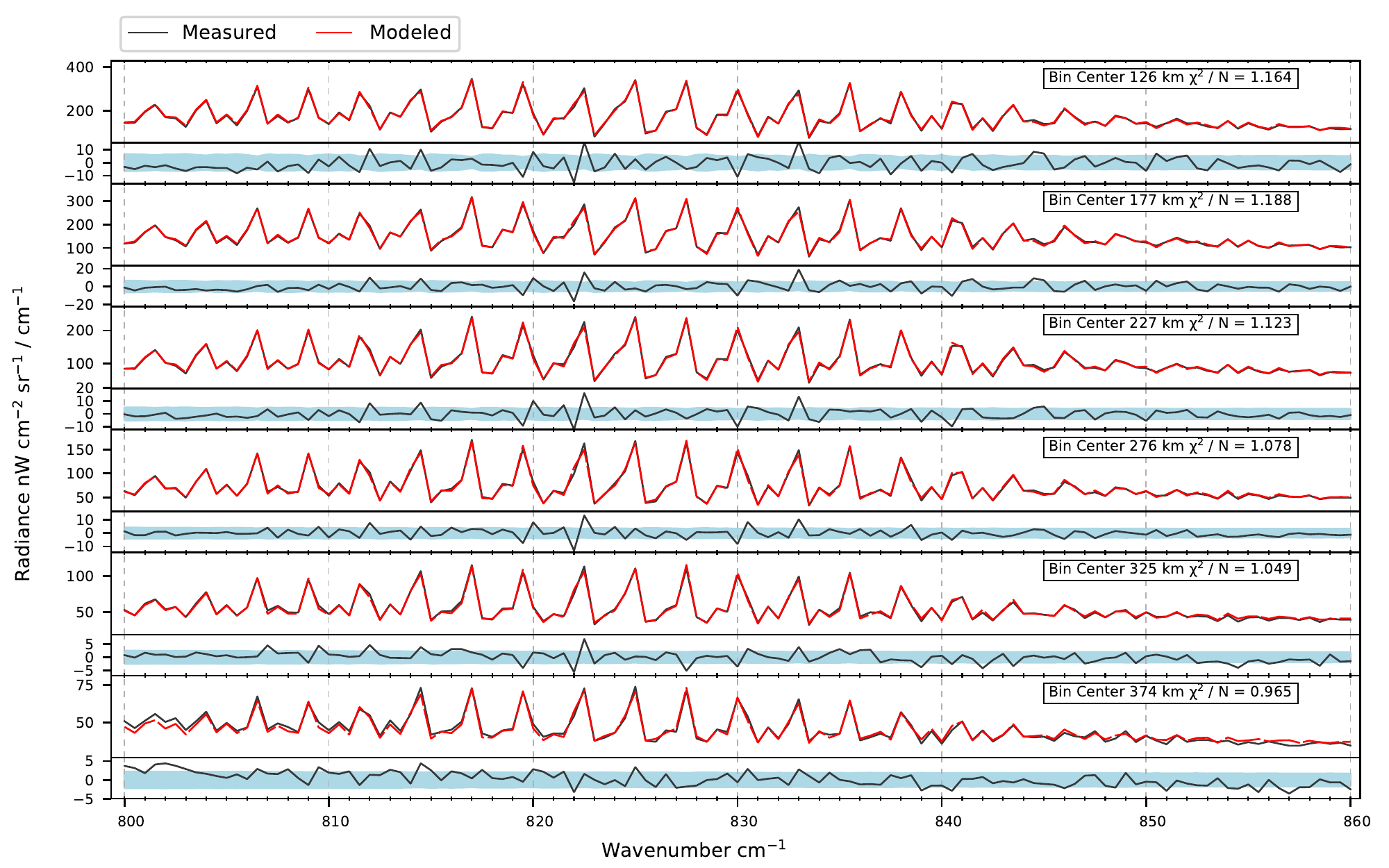}
		\caption{The same layout at as Fig. 3.  The $\nu_{12}$ band is modeled well.  The few peaks seen in the residual are from electrical interference in the instrument.  \label{fig:fp3fit}}
	\end{figure}

\begin{figure}[h]
		\centering
		\includegraphics[width = 0.5 \columnwidth]{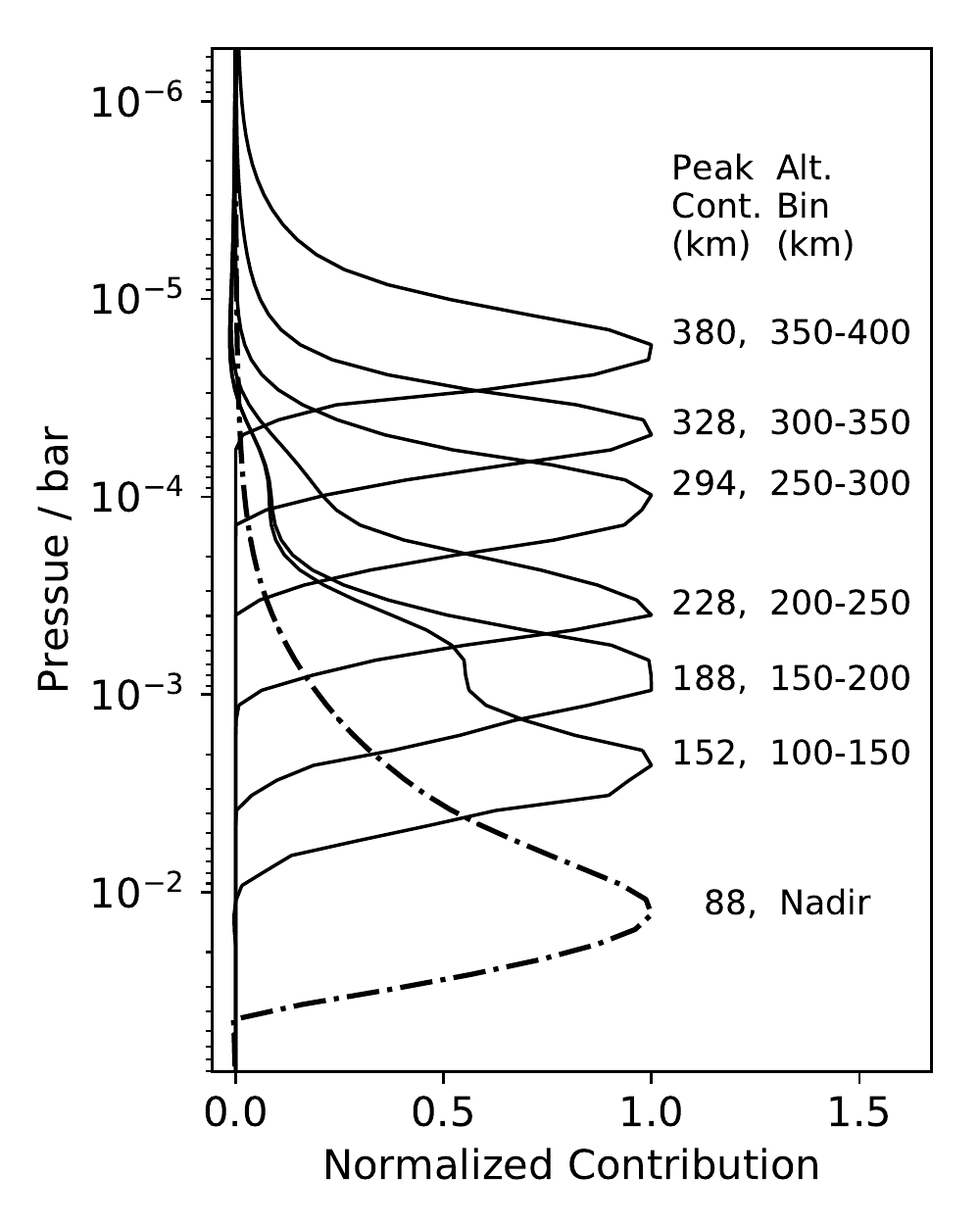}
		\caption{Normalized contribution functions at 289 cm for FP1 (dot-dashed) and 822 cm for FP3 (solid).  The contribution function from FP1 clearly peaks more than 60 km deeper in the atmosphere than the altitude bin centered at 125 km.  The contribution function can achieve a maximum value outside of the bounds of an altitude bin if the spectrum becomes optically saturated before the model `ray' looking through the atmosphere reaches the tangent altitude.  Though this band of ethane is optically thinner than the bands in FP4, a second source of emission can be seen originating from about 188 km in the lowest altitude bin being modeled, an example of the importance of the FIR retrieval.  \label{fig:fp3_cf}}
	\end{figure}

	The fit of the model to the data for FP3 is shown in Fig.~\ref{fig:fp3fit}, and is generally very good.  The few sharp spikes in the residual are due to electrical interference from the instrument, described by \cite{chan:15} and \cite{jennings:17}.  The contribution function for each altitude bin is plotted in Fig.~\ref{fig:fp3_cf}, along with the contribution function from our FP1 analysis, and shows the data are sensitive to the region between 150 and 380 km.  The retrieved vertical profile from this fit is shown in Fig.~\ref{fig:measured}, and is consistent with previous measurements from \cite{vinatier:2007, vinatier:2015}.
	
	\subsection{FP4}
	
\begin{figure}[h]
		\includegraphics[width = \columnwidth]{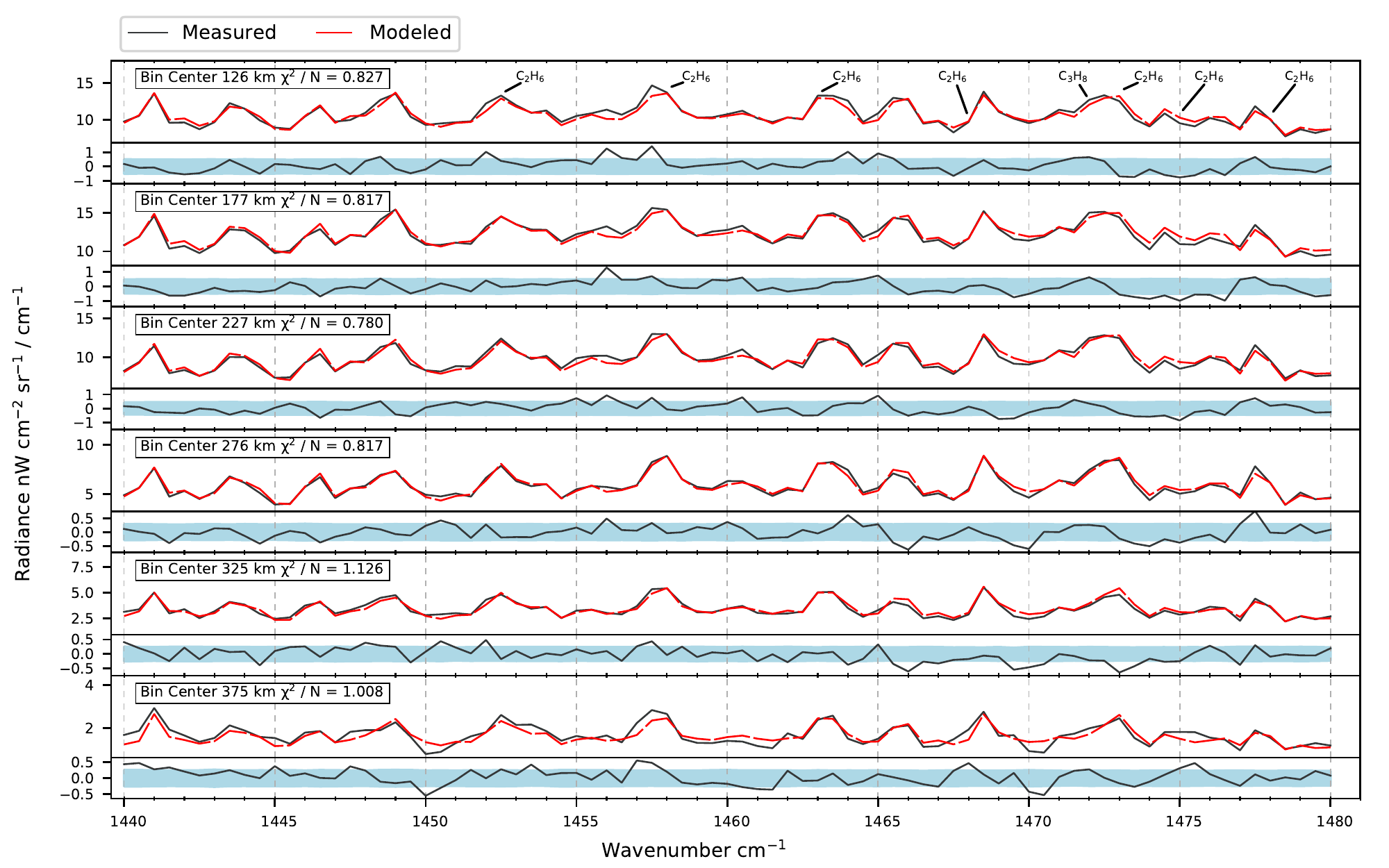}
		\caption{The same layout as Fig. 3.  The $\nu_{8}$ band is more challenging to model than the $\nu_{12}$ band because of the underlying propane band and aliasing from higher wavenumbers.  The propane band centered at 1472 cm$^{-1}$ may contribute to the slight discrepancy between measurements derived from FP3 and FP4 spectra. \label{fig:fp4fit}}
	\end{figure}

\begin{figure}[h]
		\centering
		\includegraphics[width = 0.5 \columnwidth]{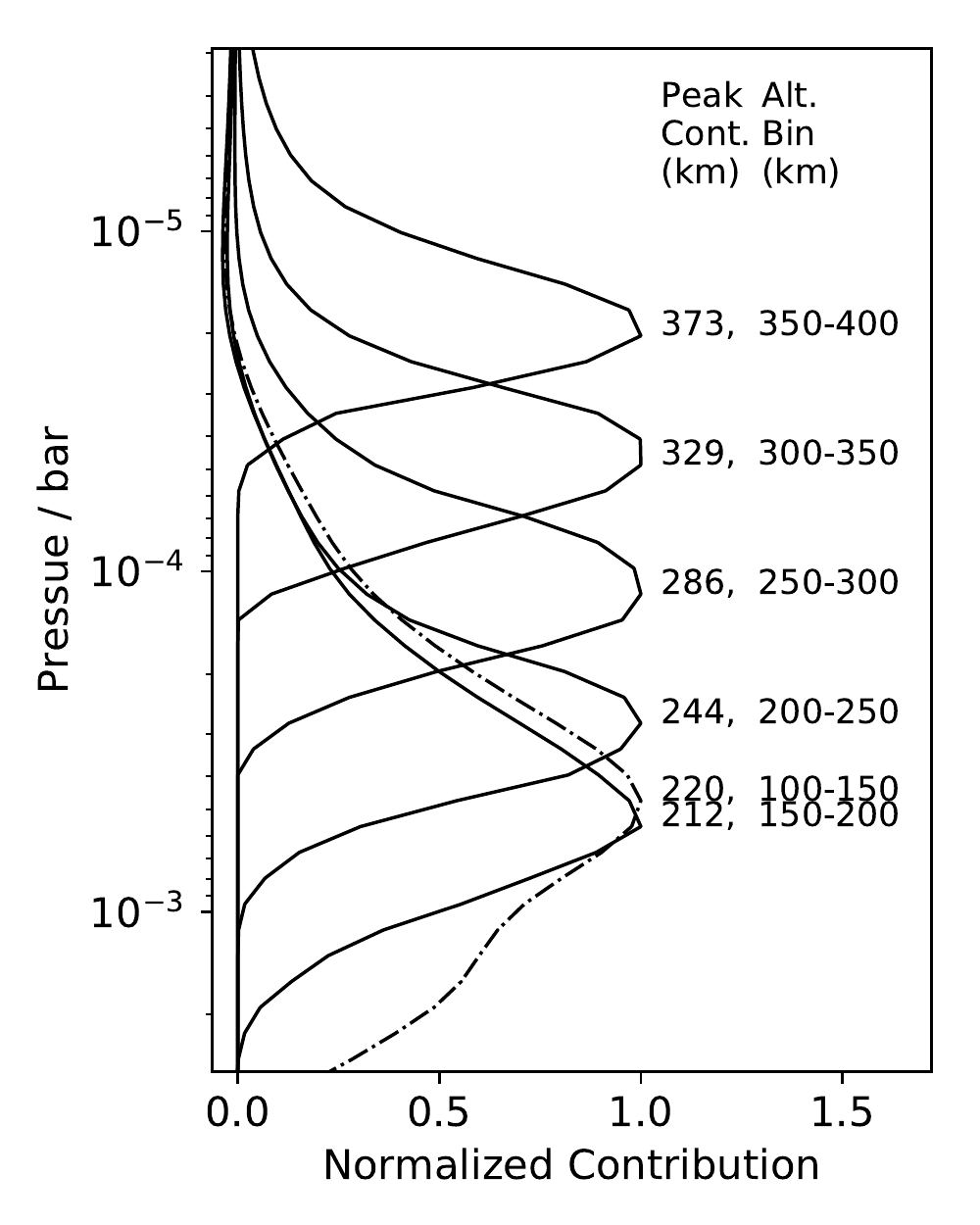}
		\caption{The contribution functions from the $\nu_{8}$ band at 1458 cm$^{-1}$ achieve maxima very high in the stratosphere due to the high opacity of the molecule at these wavelengths.  This is especially noticeable in the contribution function for the altitude bin (shown as dot-dashed for clarity and for which the greatest source of emission in this spectral window is 220 km, much higher than the tangent altitude of the observation of 125 km).  \label{fig:fp4_cf}}
	\end{figure}

	Modeling the spectral window from 1400 - 1500 cm$^{-1}$ is challenging due to the number of broad features contributed by ethane, propane, methane, and other potentially undetected trace species such as butane.  Additionally, aliasing from higher wavenumbers may affect the quality of the data in this region.  The fit of our synthetic spectra to the observations is shown in Fig.~\ref{fig:fp4fit}.  The contribution functions are plotted in Fig.~\ref{fig:fp4_cf}, which shows that even though we use the same altitude binning scheme on the limb as in our FP3 analysis, we are not sensitive to the lower altitudes.  The $\nu_{8}$ band of ethane is thus optically thicker, preventing this analysis from probing lower than 200 km.  In higher altitude bins, the sensitivity of the $\nu_{8}$ band at this resolution is comparable to the sensitivity of the $\nu_{12}$ band in FP3.  The retrieved profile from this model is shown in Fig.~\ref{fig:measured}, and is comparable to our measurements made from FP3 spectra.
	
	\section{Discussion}
		
	\subsection{Comparison between Focal Planes}
	
	\begin{table}
		\footnotesize
		\centering
		\begin{tabular}{ccc|cccc|cccc}
			\multicolumn{3}{c}{FP1}&\multicolumn{4}{c}{FP3} & \multicolumn{4}{c}{FP4}\\
			\hline
			Spectra	 &Sens. Alt.&VMR			 			&Alt. Bin	&	Spectra	&	Sens. Alt.	&	VMR			 &	Alt. Bin	 &Spectra&	Sens. Alt.		&VMR	\\
					 &km		&$\times$10$^{-5}$			&km			&			&km		&$\times$10$^{-5}$&km		&		&	km		&$\times$10$^{-5}$\\
			6684&	88 &1.0$\pm$0.4&100 - 150&	972		&152	&	1.21$\pm$0.14&	100 - 150&	1069 &	220&	1.45$\pm$0.19\\			
			&&&150 - 200&	1224	&188	&	1.53$\pm$0.14&	150 - 200&	1331 &	212&	1.42$\pm$0.19\\
			&&&200 - 250&	1406	&228	&	1.62$\pm$0.14&	200 - 250&	1555 &	244&	1.58$\pm$0.21\\	
			&&&250 - 300&	1341	&294	&	1.78$\pm$0.14&	250 - 300&	1491 &	286&	1.75$\pm$0.19\\
			&&&300 - 350&	810		&328	&	1.83$\pm$0.13&	300 - 350&	966  &	329&	1.94$\pm$0.21\\
			&&&350 - 400&	699		&380	&	2.08$\pm$0.17&	350 - 400&	760  &	373&	2.41$\pm$0.31\\
			
		\end{tabular}
		\caption{A summary of information for each altitude bin modeled.  From left to right, the columns are: the altitude bin used in the average, the number of spectra included in the average, the altitude of peak contribution for C$_{2}$H$_{6}$ at 822 cm$^{-1}$ in FP3 (left) and 1458 cm$^{-1}$ in FP4 (right), and the retrieved C$_{2}$H$_{6}$ abundance and uncertainty for each altitude bin. \label{tab:obs}}
	\end{table}
	
	The measurements made with FP3 span the region from 152 km to 380 km.  The measurements made with FP4 span the region from 212 km to 373 km.  The higher opacity of the $\nu_{8}$ band of C$_{2}$H$_{6}$ prevent limb sounding from probing the lower altitudes visible in the $\nu_{12}$ band.  In the region from ~200 km to ~400 km, however, the measured C$_{2}$H$_{6}$ abundances from each band are comparable.  These data are shown in Table \ref{tab:obs}.
	
	Compared to measurements made from the $\nu_{12}$ and $\nu_{8}$ bands of ethane in FP3 and FP4, our modeling of the $\nu_{4}$ band has probed about 50 km deeper in Titan's stratosphere.  The contribution functions of the lowest altitude bin in our FP3 and FP4 analyses achieve their maxima at altitudes of 152 km in FP3 and 212 km in FP4, 62 km and 124 km higher than the most sensitive altitude from our FP1 modeling.  The mixing ratio measured at 152 km in our FP3 analysis is (1.21$\pm$0.14)$\times$10$^{-5}$, overlapping the lower measurement at 88 km.

	\subsection{Comparison to Previous Measurements}
	
\begin{figure}[h]
		\includegraphics[width = \columnwidth]{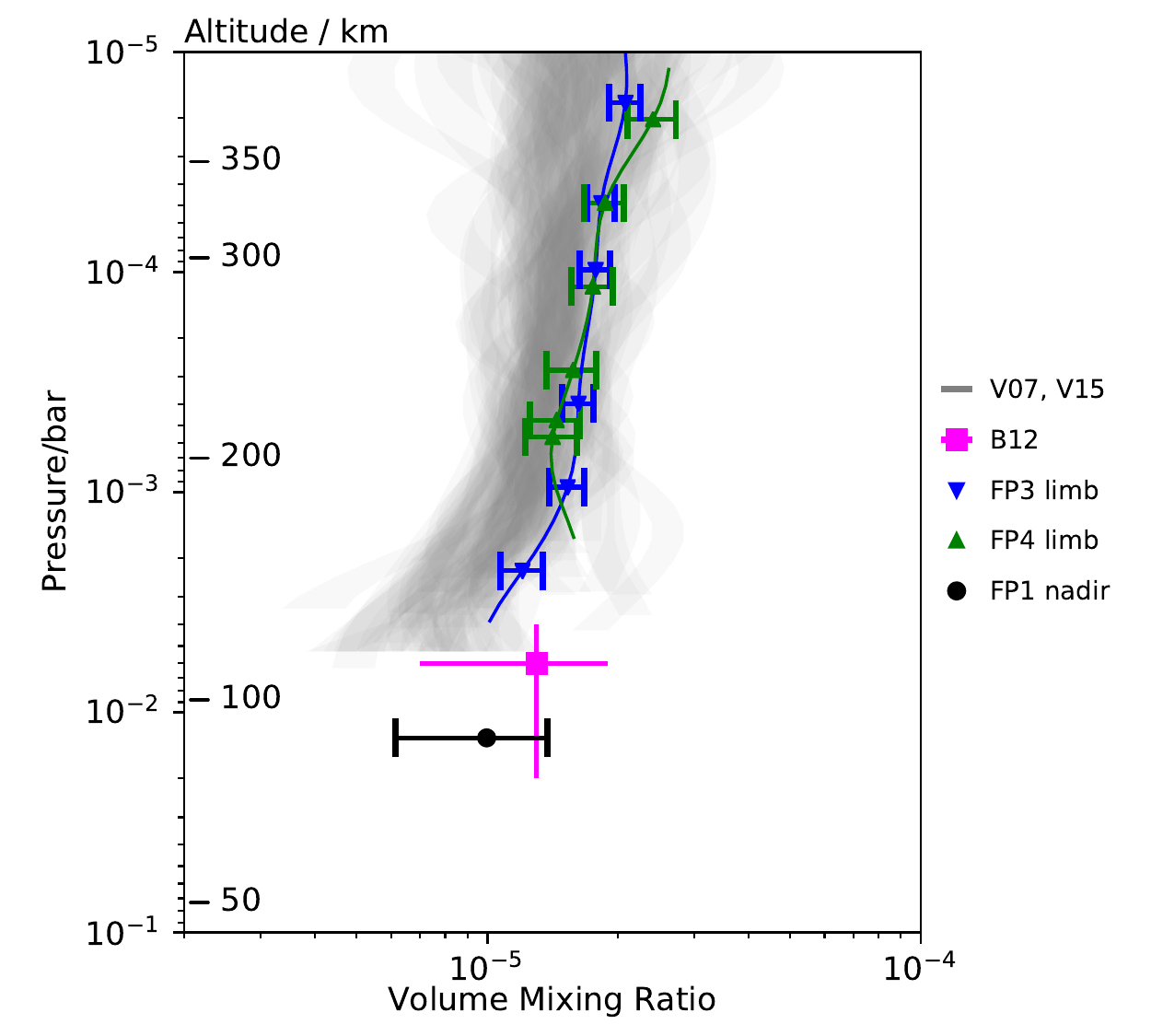}
		\caption{The abundance measured from our FP1 analysis (black), compared to profiles measured from FP3 (blue) and FP4 (green) analyses.  Our vertically resolved profiles are comparable to measurements made in \cite{vinatier:2007, vinatier:2015}, V07 and V15 in the figure.  Nadir measurements from \cite{bampasidis:2012} (pink), B12 in the figure) probe to a slightly higher altitude than that of our measurements. \label{fig:measured}}
	\end{figure}

	The measurements we made that incorporate FP3 and FP4 data are comparable to measurements from previous studies, including \cite{vinatier:2007, vinatier:2015}, \cite{coustenis:10}, and \cite{bampasidis:2012}, (Fig.~\ref{fig:measured}).  Both our measurements and the previously published measurements indicate that the mixing ratio of ethane is slowly increasing with altitude from 100 km through 400 km in Titan's stratosphere.
	
	Searching through previous literature, the ethane measurements with the deepest sounding in Titan's stratosphere are made in \cite{coustenis:10} and \cite{bampasidis:2012}, who retrieve ethane abundance from CIRS FP3 nadir spectra.  The contribution function reported in \cite{coustenis:10} achieves a maximum at 7 mbar, roughly 100 km.  We assume that \cite{bampasidis:2012} are sensitive to the same altitude, as both groups use the same method and similar datasets.  The span of mixing ratios reported in \cite{bampasidis:2012} are plotted in Fig.~\ref{fig:measured} for comparison with our measurements.  The vertical error on this data point is the FWHM of the contribution function for ethane reported in \cite{coustenis:10}, and the horizontal error is the span of ethane measurements, including error, reported in \cite{bampasidis:2012}.
	
	\subsection{Comparison to Photochemical Models}
	
	\begin{figure}[h]
		\includegraphics[width = \columnwidth]{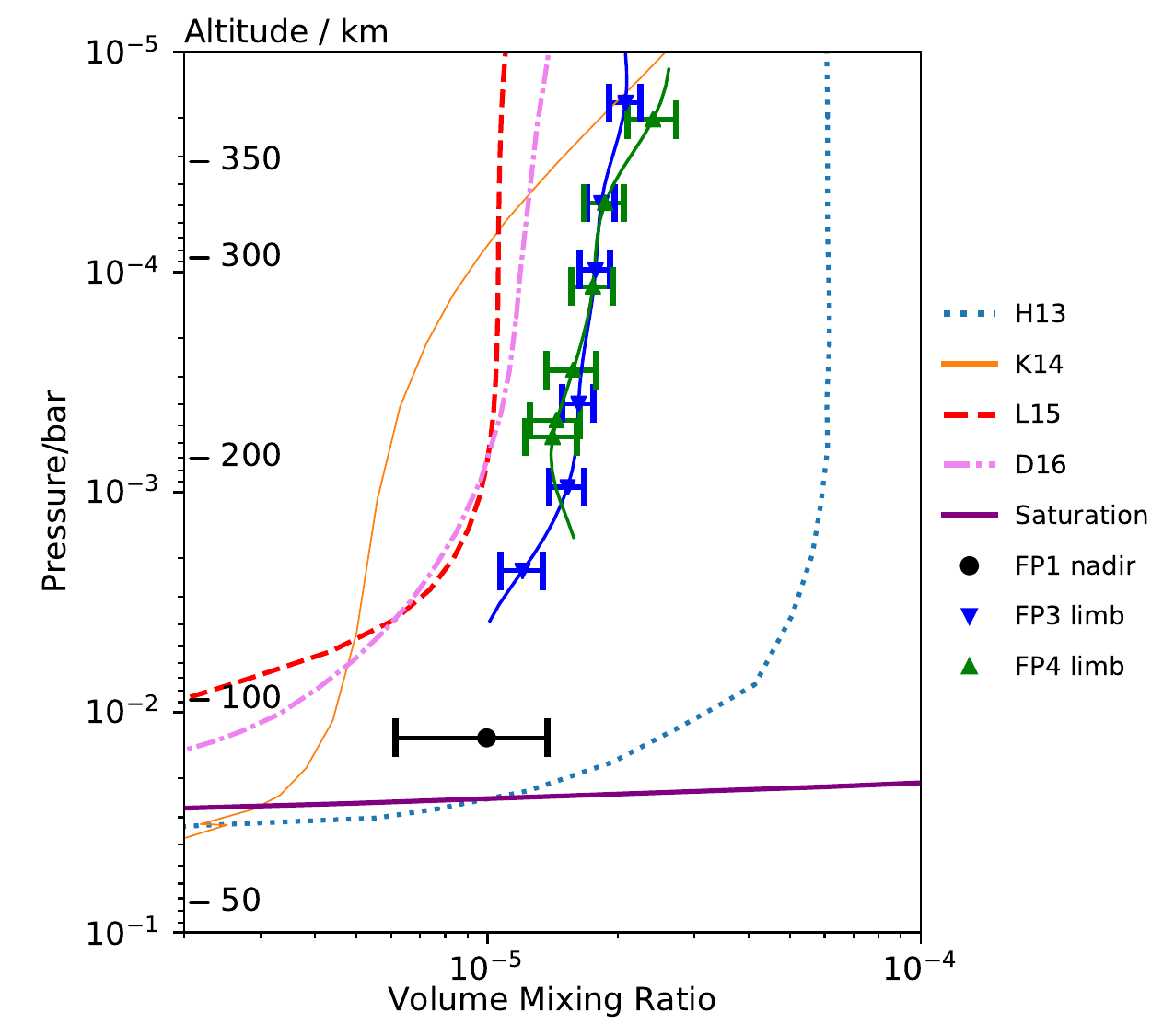}
		\caption{
			Our measurements at 88 km made from FP1 data and above 100 km for FP3 and FP4 are compared with the predictions from photochemical models.  H13 is from \cite{hebrard:2013}, K14 is from \cite{kras:2014}, L15 is from \cite{loison:15}, D16 is from \cite{dobrijevic:2016}.  The purple solid line is the saturation abundance of ethane at the temperatures and pressures included in our model.  When compared with measurements from mid-infrared observations, our analysis shows that ethane is about as abundant at 88 km as it is between 100 km and 400 km.  In contrast, photochemical models predict ethane to be depleted in this region. \label{fig:model}}
	\end{figure}
	
	To evaluate the accuracy of photochemical models, we compare our measurement at 88 km to predictions from \cite{hebrard:2013}, \cite{kras:2014}, \cite{li:2015}, and \cite{dobrijevic:2016}, plotted in Fig.~\ref{fig:model}.  With the exception of the model presented in \cite{hebrard:2013}, which overestimates ethane throughout the stratosphere by a factor of 6, the models we compare to predict a mixing ratio about an order of magnitude lower than what we measure at 88 km.  All of the models predict that ethane is depleted in this region compared to altitudes above 100 km.  Our measurements show that at 88 km, ethane is nearly as abundant as higher in the stratosphere.  The discrepancy could be caused by overestimating depletion mechanisms, underestimating production mechanisms, or inaccurate eddy diffusion coefficients.  We come to a similar conclusion when comparing to \cite{vuitton:18}, who predict a lower concentration of C$_{2}$H$_{6}$ throughout the stratosphere relative to our measurements.
	
	\section{Conclusion}
	Ethane is a prominent component of Titan's atmosphere, surface, and potentially sub-surface.  The molecule is formed high in the atmosphere and can be transported deep into the stratosphere where it may condense into clouds or precipitate onto the surface.  Ethane is also expected to be a component of Titan's lakes where it may interact with the troposphere and surface of Titan.  The molecule may also interact with the surface materials on Titan, potentially opening a channel to transport ethane into Titan's subsurface.  To fully understand the role of ethane on Titan, we must have accurate measurements of the molecule throughout the atmosphere.
	
	In this paper, we have measured ethane in Titan's stratosphere by modeling its $\nu_{4}$ band observed in CIRS far-infrared spectra.  This marks the first time that this band has been used to measure the abundance of ethane in a planetary atmosphere.  Using this optically thin band has enabled us to probe a lower altitude than permitted by the optically thicker mid-infrared $\nu_{12}$ and $\nu_{8}$ bands.
	
	We have also modeled the $\nu_{8}$ band of ethane, showing that it can be a useful tool in measuring ethane in planetary atmospheres.  Future observations of planetary atmospheres may be able to make use of our results by observing both the $\nu_{8}$ band of C$_{2}$H$_{6}$ and the nearby $\nu_{4}$ band of CH$_{4}$, frequently used as an atmosphere thermometer, simultaneously.
	
	\section{Acknowledgments}
	N.A.L., C.A.N., R.K.A., and F.M.F. were supported by the NASA Cassini Project for the research work reported in this paper.  M.S., N.A.T., and P.G.J.I. were funded by the UK Science and Technology Facilities Council.
	
	\bibliography{ethane}

\end{document}